\pdfoutput=1
\RequirePackage{ifpdf}
\ifpdf %We are running pdfTeX in pdf mode
\documentclass[pdftex]{sigma}
\else
\documentclass{sigma}
\fi

\usepackage[all]{xy}

\begin{document}

\newcommand{\arXivNumber}{1501.06601}

\allowdisplaybreaks

\renewcommand{\PaperNumber}{038}

\FirstPageHeading

\ShortArticleName{Invariant Classif\/ication and Limits of Maximally Superintegrable Systems in 3D}

\ArticleName{Invariant Classif\/ication and Limits\\
of Maximally Superintegrable Systems in 3D}

\Author{Joshua J.~CAPEL~$^\dag$, Jonathan M.~KRESS~$^\dag$ and Sarah POST~$^\ddag$}

\AuthorNameForHeading{J.J.~Capel, J.M.~Kress and S.~Post}

\Address{$^\dag$~Department of Mathematics, University of New South Wales, Sydney, Australia}
\EmailD{\href{mailto:j.capel@unsw.edu.au}{j.capel@unsw.edu.au}, \href{mailto:j.kress@unsw.edu.au}{j.kress@unsw.edu.au}}

\Address{$^\ddag$ Department of Mathematics, University of Hawai`i at M\=anoa, Honolulu, HI, 96822, USA}
\EmailD{\href{mailto:spost@hawaii.edu}{spost@hawaii.edu}}

\ArticleDates{Received February 03, 2015, in f\/inal form April 21, 2015; Published online May 08, 2015}

\Abstract{The invariant classif\/ication of superintegrable systems is reviewed and utilized to construct singular limits
between the systems.
It is shown, by construction, that all superintegrable systems on conformally f\/lat, 3D complex Riemannian manifolds can
be obtained from singular limits of a~generic system on the sphere.
By using the invariant classif\/ication, the limits are geometrically motivated in terms of transformations of roots of
the classifying polynomials.}

\Classification{33C45; 33D45; 33D80; 81R05; 81R12}

\section{Introduction}

The discovery and classif\/ication of superintegrable systems is currently undergoing signif\/icant activity,
see~\cite{superreview} and citations within.
Superintegrable systems are dynamical systems, for example classical or quantum Hamiltonian systems, that have more
conserved quantities than degrees of freedom.
One of the many active areas of research in superintegrable systems is in their classif\/ication.
On this front, there has been signif\/icant progress on maximally superintegrable systems in 2 and 3-D complex Riemannian
manifolds with constants of the motion that are at most second-order in the momenta, a~necessary condition for
multi-separability.
In two dimensions, all such systems have been classif\/ied~\cite{KKM20061}.
Somewhat surprisingly, all such systems can be obtained through singular limits of a~``generic'' system def\/ined on the
sphere.
This fact was f\/irst recognized by B\^{o}cher~\cite{Bocher} in his search for metrics that admit the maximal number of
second-order Killing tensors.
Again in the 2D case, the limits between such systems were recently given explicitly in~\cite{KMPost13}, where they were
shown to induce contractions on the associated symmetry algebras as well as limits in the representations of such
algebras.
These representations are given in terms of classical hypergeometric orthogonal polynomials and the limits of the
superintegrable systems correspond to the limits within the Askey scheme.

The aim of this paper is to give the corresponding limits of superintegrable systems in 3D.
We will show explicitly that all such systems are limiting cases of a~``generic'' system def\/ined on $\mathbb{S}^3$ and
give the coordinate transformation that generates each limit.
In order to describe the limits we discuss the classif\/ication theory of 3D second-order superintegrable
systems~\cite{CapelThesis, CapelKress2014}.
The classif\/ication makes use of classical invariant theory and associates to each superintegrable system a~7-dimensional
representation of the conformal group.
By studying the action of the conformal group on this representation, we can identify 10 equivalence classes each of
which corresponds to a~class of superintegrable systems.
From the invariant classif\/ication, the necessary limits are abstractly motivated as opposed to the constructions
of~\cite{KMPost13}, which are given ad hoc.

The structure of the paper is as follows.
We begin in Section~\ref{Section2} with an overview of the results in~\cite{CapelThesis, CapelKress2014} giving the invariant  classif\/ication of nondegenerate second-order superintegrable systems in 3D.
Section~\ref{Section3} contains the results of the paper; namely we prove that all such superintegrable systems are limits of
a~``generic'' system by demonstrating the appropriate limits.
Section~\ref{Section4} contains a~brief discussion of the results and future applications of this research.

\section{Invariant classif\/ication of superintegrable systems}\label{Section2}

In this section, we describe the invariant classif\/ication of nondegenerate second-order superintegrable systems in 3D,
where the metric is assumed to be conformally f\/lat.
The dynamics are given by the following classical Hamiltonian on 6-dimensional phase space $({\bf p,x})$
\begin{gather*}
\mathcal{H}({\bf p, x})=\sum\limits_{i,j=1}^3g^{ij}({\bf x})p_ip_j+V({\bf x}).
\end{gather*}
The conformally f\/lat assumption implies that the metric tensor takes the form $g^{ij}=\lambda({\bf x})\delta_{j}^i$.
Any function on the phase space $\mathcal{A}({\bf p,x})$ will transform in time as
\begin{gather*}
\frac{d \mathcal{A}}{dt}=\{\mathcal{A}, \mathcal{H}\} +\frac{\partial A}{\partial t},
\end{gather*}
where $\{\,, \,\}$ is the standard Poisson bracket
\begin{gather*}
\{\mathcal{A}, \mathcal{B}\}\equiv \sum\limits_{i=1}^3 \frac{\partial \mathcal{A}}{\partial x_i}\frac{\partial
\mathcal{B}}{\partial p_i}-\frac{\partial \mathcal{B}}{\partial x_i}\frac{\partial \mathcal{A}}{\partial p_i}.
\end{gather*}
Thus, a~function on phase-space will be invariant under the Hamiltonian f\/low if and only if it Poisson commutes with
$\mathcal{H}$, ($\{\mathcal{A}, \mathcal{H}\}=0$) and is referred to as a~constant of the motion.
Conversely, the Hamiltonian is also invariant under f\/low generated by the constant of the motion and so the function
${\mathcal {A}}$ is also often referred to as a~symmetry of the Hamiltonian.

A Hamiltonian on $2n$-dimensional phase space is said to be Liouville integrable if there exist~$n$ mutually commuting
constants of the motion that are functionally independent.
We say that a~Hamiltonian is superintegrable if there exist more than~$n$ constants of the motion and maximally
superintegrable if there exist $2n-1$ such symmetries.
Not all of the symmetries can mutually commute and so we drop the requirement that the integrals commute among
themselves.
There is a~close connection between superintegrability and noncommutative integrability, notably that maximally
superintegrable systems are noncommutatively integrable.
Indeed, the celebrated result of Nekhoroshev~\cite{nekhoroshev1972} concerning foliations of phase space for
noncommutative integrable systems includes, as a~special case, the result that, generically, bounded trajectories of superintegrable
systems are closed.

We would like to classify Hamiltonians ${\mathcal {H}}$ on a~conformally f\/lat 3D manifold with 5 constants of motion
that are second-order in the momentum, i.e., that there exist 5 (including the Hamiltonian) functionally independent
constants of the motion
\begin{gather*}
{\mathcal {A}}=\sum\limits_{i,j=1}^3a^{ij}({\bf x}) p_ip_j+W({\bf x}),
\qquad
a^{ij}=a^{ji},
\end{gather*}
such that
\begin{gather*}
\{{\mathcal {A}}, {\mathcal {H}} \}=0.
\end{gather*}
If such symmetries exist then a~direct computation of the Poisson commutators leads to the following set of dif\/ferential
equations: the Killing equations, which require the second-order part of ${\mathcal {A}} $ to be a~Killing tensor on the
manifold, and the Bertrand--Darboux (BD) equations.
The BD equations are the compatibility conditions for the dif\/ferential equations for $W({\bf x})$ and comprises a~set
of four linear second-order PDEs for the potential $V({\bf x})$, one for each integral that is not the Hamiltonian.
Combining the equations into vector form gives
\begin{gather*}
%\label{MBD}
B\big(a^{ij}\big) \left(
\begin{matrix} V_{,22}-V_{,11}
\\
V_{,33} -V_{,11}
\\
V_{,12}
\\
V_{,13}
\\
V_{,23}
\end{matrix}
\right)=C\big(\lambda, \lambda_{,k}, a^{ij}, a^{ij}_{,k}\big) \left(
\begin{matrix} V_{,1}
\\
V_{,2}
\\
V_{,3}
\end{matrix}
\right),
\end{gather*}
where the subscript $f_{,k}$ denotes the partial derivative with respect to $x_k$ and~$B$ and~$C$ are matrix functions
of the given arguments of dimensions $12\times 5$ and $12\times 3$, respectively.
Suppose in addition that the constants of the motion are functionally, linearly independent.
This implies that the BD equations are of rank 5 and it is possible to solve for second-order derivatives of~$V$ as in
\begin{gather}
\label{reducedBD}
\left(
\begin{matrix} V_{,11}
\\
V_{,22}
\\
V_{,33}
\\
V_{,12}
\\
V_{,13}
\\
V_{,23}
\end{matrix}
\right)= \left(
\begin{matrix}
1& -4S^1-R^{12}_2-R^{13}_3& 2S^2+R^{12}_1&2S^3+R^{13}_1
\\
1& 2S^1+R^{12}_2& -4S^2-R^{12}_1-R^{23}_3& 2S^3+R^{23}_2
\\
1&2S^1+R^{13}_3& 2S^2+R^{23}_3& -4S^{3}-R^{13}_1-R^{23}_2
\\
0&R^{12}_1-3S^2& R^{12}_2-3S^1& Q^{123}
\\
0&R^{13}_1-3S^3& Q^{123}&R^{13}_3-3S^1
\\
0&Q^{123}& R^{23}_2-3S^3& R^{23}_3-3S^2
\end{matrix}
\right)\vec{v},
\\
\vec{v}\equiv \left(
\begin{matrix}V_{,ee}
\\
V_{,1}
\\
V_{,2}
\\
V_{,3}
\end{matrix}
\right),
\qquad
V_{,ee}=\frac13(V_{,11}+V_{,22}+V_{,33}).
\nonumber
\end{gather}
Here $V_{,ee}$ is not a~second derivative, but a~symmetry adapted variable.
The use of~$V_{,ee}$ introduces redundancy in the equations (6 equations instead of 5), which is preserved for the sake
of symmetry in the variables.
If the compatibility conditions of~\eqref{reducedBD} are assumed to hold identically
then the potential depends on 5~parameters, the values of $V_{,ee}$, $V_{,1}$, $V_{,2}$, $V_{,3}$
and~$V$ at a~generic point, the last point coming from a~trivial additive parameter.
Such a~system is said to be {\it non-degenerate}. For a~non-degenerate system, the potential is uniquely determined~by
the value of the matrix in~\eqref{reducedBD} at a~given generic point ${\bf x}_0$.
The values of $V({\bf x}_0)$, $V_{,1}({\bf x}_0)$, $V_{,2}({\bf x}_0)$, $V_{,3}({\bf x}_0)$ and $V_{,11}({\bf x}_0)$ are
the parameters in the potential.
As is clear from the form of the matrix, it depends on 10 functions
\begin{gather*}
\big\{Q^{123}, S^1, S^2, S^3, R^{12}_1, R^{12}_2, R^{13}_1, R^{13}_3, R^{23}_2, R^{23}_3\big\}\equiv {\bf \{Q, S, R\}},
\end{gather*}
which are assumed to satisfy the compatibility conditions for~\eqref{reducedBD} identically.
The compatibility conditions of~\eqref{reducedBD} are f\/irst-order equations in the variables whose compatibility
conditions are themselves identically satisf\/ied.
Therefore, the values of functions ${\bf \{Q, S, R\}}$ at a~generic point~${\bf x}_0$ are enough to uniquely determine
the functions themselves along with a~superintegrable system.

\begin{example}
The harmonic oscillator potential
\begin{gather*}
V_{OO}=\frac{a}{2}\big(x_1^2+x_2^2+x_3^2\big)+bx_1+c x_2+dx_3+e
\end{gather*}
is a~non-degenerate second-order superintegrable system.
The BD equations for the potential are
\begin{gather*}
\left(
\begin{matrix} V_{,11}
\\
V_{,22}
\\
V_{,33}
\\
V_{,12}
\\
V_{,13}
\\
V_{,23}
\end{matrix}
\right)= \left(
\begin{matrix} 1& 0& 0&0
\\
1& 0& 0& 0
\\
1&0& 0& 0
\\
0&0& 0&0
\\
0&0& 0 &0
\\
0&0& 0& 0
\end{matrix}
\right)\left(
\begin{matrix}V_{,ee}
\\
V_{,1}
\\
V_{,2}
\\
V_{,3}
\end{matrix}
\right).
\end{gather*}
Taking the origin as a~generic point, the coef\/f\/icients are related to~$V$ as $a=V_{,ee}({\bf 0})$, $b=V_{,1}({\bf 0})$,
$c=V_{,2}({\bf 0})$, $d=V_{,3}({\bf 0})$, and $e=V({\bf 0})$.
\end{example}

\subsection{Equivalence classes}

The next step in the classif\/ication is to identify the appropriate equivalence classes.
Clearly, we would like to consider two potentials as equivalent if they are related by a~change of parameters.
This is accomplished in the previous section by making a~canonical choice of parameters as the value of $\vec{v}({\bf x}_0)$.
We would also like systems to be equivalent if they are related by translations in the position variables; such
a~translation corresponds to moving the regular point.
We would also like systems to be equivalent if they are related by, possibly complex, rotations.
This condition will be studied extensively in this section.
Finally, we would like the equivalence classes to include systems that are related via the St\"ackel transform or
coupling constant metamorphosis~\cite{HGDR, KKM20052, KMP10, Post20111}.
That is, suppose a~Hamiltonian can be expressed as ${\mathcal {H}}={\mathcal {H}}_0+\alpha U$ then the St\"ackel
transformed Hamiltonians $\hat{{\mathcal {H}}}=\frac{1}{U}{\mathcal {H}} $ will also be superintegrable, perhaps on
a~dif\/ferent conformally f\/lat manifold.
Under this transformation, classical trajectories and quantum wave functions as well as the corresponding symmetry
algebras are essentially preserved, up to a~change of parameters.
Thus, we would like to consider two such systems as being equivalent.
An analogous classif\/ication of equivalence classes for superintegrable systems in 2D has already been
performed~\cite{Kress2007}.

In order to consider St\"ackel equivalent systems, we focus our attention on {\it conformally} superintegrable
systems~\cite{KKMP2011}.
For conformal superintegrable systems, the Hamiltonian is identi\-cal\-ly~0 on trajectories.
This can be accomplished mathematically by including the trivial added parameter to the system.
Thus, along a~trajectory the Hamilton--Jacobi
equation is expressed as ${\mathcal {H}}-E=0$ but with ${\mathcal {H}} -E$
being the Hamiltonian.
A~function on phase space will then be a~constant of the motion whenever its Poisson bracket is given~by
\begin{gather*}
\{\mathcal{A}, {\mathcal {H}}\}=f({\bf p,x}) {\mathcal {H}}.
\end{gather*}
It is a~direct computation to verify (see, e.g., \cite[Lemma 4.1.4]{CapelThesis}) that a~conformal integral of the
motion $\mathcal{A}$ of ${\mathcal {H}}$ will be a~conformal integral of the scaled Hamiltonian $U({\bf x}){\mathcal{H}}$.
Thus it is possible to transform any conformal superintegrable Hamiltonian on conformally f\/lat space to a~conformally
superintegrable Hamiltonian on Euclidean space.
Furthermore, since the action of the St\"ackel transform corresponds to multiplying the Hamiltonian by a~function, it is
clear that two Hamiltonians equivalent under this action will be equivalent conformal superintegrable systems, from the
perspective of their symmetry algebras.
\begin{example}
The Hamiltonian for the simple Harmonic oscillator ${\bf p}^2-\omega^2 {\bf x}^2$ corresponds to the following conformal
Hamiltonian
\begin{gather*}
{\mathcal {H}}={\bf p}^2-\omega^2 {\bf x}^2-E.
\end{gather*}
Dividing by ${\bf x}^2$ gives
\begin{gather*}
\widehat{{\mathcal {H}}}=\frac1{{\bf x}^2}{\bf p}^2-\omega^2-\frac{E}{{\bf x}^2}.
\end{gather*}
Clearly, trajectories that satisfy ${\mathcal {H}}=0$ will also satisfy $\widehat{{\mathcal {H}}}=0$, however the
``energy'' of the new Hamiltonian is given by the parameter~$\omega^2$.
\end{example}
Conversely, any conformally superintegrable system that depends linearly on at least one arbitrary constant can be
transformed, see, e.g., \cite[Theorem 4.1.8]{CapelThesis}, into a~superintegrable system by coupling constant
metamorphosis.
Therefore, classifying superintegrable systems on conformally f\/lat manifolds up to St\"ackel transform is equivalent to
classifying {\it conformally superintegrable systems} on Euclidean space.
The determining equations for an integral of a~conformal system are similar to~\eqref{reducedBD} except that the
additive constant is no longer trivial but instead essential for specifying the Hamiltonian.
The corresponding structure equations are then
\begin{gather*}
%\label{extendedBD}
\left(
\begin{matrix}
V_{11}
\\
V_{,22}
\\
V_{,33}
\\
V_{,12}
\\
V_{,13}
\\
V_{,23}
\end{matrix}
\right)= A\left(
\begin{matrix}
V_{,ee}
\\
V_{,1}
\\
V_{,2}
\\
V_{,3}
\\
V
\end{matrix}
\right),
\end{gather*}
with
\begin{gather*}
A= \left(
\begin{matrix}
1& -4S^1-R^{12}_2-R^{13}_3& 2S^2+R^{12}_1&2S^3+R^{13}_1&A^{11}_0
\\
1& 2S^1+R^{12}_2& -4S^2-R^{12}_1-R^{23}_3& 2S^3+R^{23}_2&A^{22}_0
\\
1&2S^1+R^{13}_3& 2S^2+R^{23}_3& -4S^{3}-R^{13}_1-R^{23}_2&A^{33}_0
\\
0&R^{12}_1-3S^2& R^{12}_2-3S^1& Q^{123}&A^{12}_0
\\
0&R^{13}_1-3S^3& Q^{123}&R^{13}_3-3S^1&A^{13}_0
\\
0&Q^{123}& R^{23}_2-3S^3& R^{23}_3-3S^2&A^{23}_0
\end{matrix}
\right).
\end{gather*}
The functions $A^{jk}_0$ are quadratic functions in the $\{{\bf Q, S, R}\}$ given in~\cite{CapelThesis}.

\subsection{Action of the conformal group}

So far, we have determined that superintegrable systems are uniquely determined by the value of the 10 functions $\{{\bf Q, R, S}\}$
at a~regular point.
We are interested in determining equivalence classes of systems that are stable under the action of the conformal group
so we look at the induced action of this group on these 10 functions.
Again, the exact formulas and derivations can be found in~\cite{CapelThesis, CapelKress2014} and here we state only the
results necessary to understand the limits obtained in Section~\ref{Section3}.

A conformal change of variables can be generated by translations, which act trivially on the functions $\{{\bf Q, R, S}\}$
and inversions in the spheres of varying radii, which decompose the functions into
a~7-dimensional representation $\{{\bf Q, R}\}$ and a~3-dimensional
representation carried by $\{{\bf S}\}$.
Signif\/icantly, we can use a~group transformation to set the values of $\{{\bf S}\}$ to any desired value, described in
the following theorem.

\begin{theorem}
[re-statement of~\protect{\cite[Theorem 4.2.18]{CapelThesis}}] Given a~conformally superintegrable system with function values
$\{{\bf Q_0, R_0, S_0}\}$ at a~regular point ${\bf x}_0$, there exists a~conformal group motion that maps the function
values to $\{{\bf Q_0, R_0, 0}\}$ at the transformed regular point ${\bf \hat{x}}_0$.
\end{theorem}
This theorem allows us restrict our attention to the action of the conformal group on the 7-dimensional space
$\{{\bf Q, R}\}$.
To understand the action, we consider the continuous generators of the conformal group, namely translations, scaling,
rotations and M\"obius transformations.
As discussed before, translations act trivially on our representations.
Scaling the coordinates corresponds to a~scaling of the functions.
More interesting is the action of rotations.
To represent this action, we form weight vectors from the functions
\begin{gather*}
Y_{\pm3}  = R^{12}_{1}+\frac{1}{4} R^{23}_{3}\pm i\left(R^{12}_{2}+\frac{1}{4} R^{13}_{3}\right),
\qquad
Y_{\pm2}  = \frac{1}{4}\sqrt{6} \big(i\big(R^{13}_{1}-R^{23}_{2}\big)\mp 2 Q^{123}\big),
\\
Y_{\pm1}  = \frac{1}{4}\sqrt{15} \big(R^{23}_{3}\mp i R^{13}_{3}\big),
\qquad
Y_{0}  = -\frac{1}{2} i \sqrt{5} \big(R^{13}_{1}+R^{23}_{2}\big),
%\label{Y_representation}
\end{gather*}
so that the action of rotations are given by the standard raising and lowering operators for~$\mathfrak{so}(3)$. Using
the isomorphism between $\mathfrak{so}_3(\mathbb{C})$ and $\mathfrak{sl}_2(\mathbb{C})$, we obtain a~covariant
representation of the action via the polynomial
\begin{gather*}
q(z)=\sum\limits_{j=0}^6 (-1)^j\sqrt{\binom{6}{j}}Y_{3-j}z^{6-j}.
%\label{qcov}
\end{gather*}
Here the action of ${\rm SO}_3(\mathbb{C})$ is represented, via the isomorphism, as the standard action of ${\rm SL}_2(\mathbb{C})$
as M\"obius transformations
\begin{gather}
\label{mob}
\hat{q}(\hat{z})=(c\hat{z}+d)^6 q\left(\frac{a\hat{z}+b}{c\hat{z}+d}\right).
\end{gather}
Thus, the action of the conformal group on the functions $\{{\bf Q, R}\}$ can be represented by the action of
${\rm GL}_{2}(\mathbb{C})=\mathbb{C}\times {\rm SL}_2(\mathbb{C})$ on the polynomials $q(z)$ via scaling and~\eqref{mob}.
Furthermore, the action of M\"obius transformations on the coordinates can be represented locally as rotation combined
with a~scaling and so, if we f\/ind invariants that are closed under translation, scaling and rotations they will be
automatically closed under the entire action of the conformal group.
Finally, we have the following theorem.

\begin{theorem}
[Theorem~4.2.21 of~\cite{CapelThesis} in non-homogeneous coordinates] Given a~conformal superintegrable system and a~regular
point, there is a~local conformal transformation $($i.e., excluding translation of the regular point$)$ taking it to the
regular point of another superintegrable system if and only if the roots of the corresponding covariant polynomial at
the corresponding regular points are equivalent up to a~general linear transform.
\end{theorem}

Therefore, a~key to understanding the classif\/ication of conformal superintegrable systems is to understand the
classif\/ication of invariants of the roots of degree 6 polynomials.
Furthermore, since we would like the equivalence classes to be stable under translation of the regular point, we require
that any invariants obtained be closed under derivations.
This signif\/icantly reduces the number of equivalence classes to 10, exactly the expected number of equivalent
superintegrable systems.
For more details on the classif\/ication, we refer the reader to the original papers~\cite{CapelThesis,CapelKress2014}.

\section{Contractions}\label{Section3}

For each of the contractions, we need only record the following data.
The initial regular point $\bf{x}_0$, the rotation angles $t_1$, $t_2$, $t_3$, the scaling parameter~$c$ and the f\/inal
regular point $\bf{y}_0$.
We note that the rotations can be represented in ${\rm SL}_{2}(\mathbb{C})$ by the following matrices:
\begin{gather*}
\rho(R_3)=\left[
\begin{matrix}
e^{it_3/2}&0
\\
0& e^{-it_3/2}
\end{matrix}
\right],
\qquad
\rho(R_2)= \left[
\begin{matrix}
\cos(t_2/2) &- \sin(t_2/2)
\\
\sin(t_2/2) & \cos(t_2/2)
\end{matrix}
\right],
\\
\rho(R_1)=\left[
\begin{matrix}
\cos(t_1/2) &- i\sin(t_1/2)
\\
-i\sin(t_1/2) & \cos(t_1/2)
\end{matrix}
\right].
\end{gather*}
The scaling~$c$ can be encoded in the same manner by the matrix
\begin{gather*}
\rho(c)=\left[
\begin{matrix}
c^{-1/6} & 0
\\
0 & c^{-1/6}
\end{matrix}
\right].
\end{gather*}
In the following sections, unless otherwise mentioned, the angles are assumed to be set to 0 and the scale factor~$c$ to
be set to 1.
The action of the matrices on the polynomials are given~by
\begin{gather*}
\rho(A) \circ q(z)=\hat{q}(\hat{z}),
\qquad
\rho(A)=\left[
\begin{matrix}
a&b
\\
c&d
\end{matrix}
\right],
\end{gather*}
as in~\eqref{mob}, which transform the roots as
\begin{gather*}
%\label{rootshat}
\hat{\mathbf{a}}=\frac{d\mathbf{a}-b}{-c\mathbf{a}+a}.
\end{gather*}
To emphasize the dependence of the polynomials on the regular point, we write $q(x_1, x_2, x_3)(z)$ where necessary.

The action of the ${\rm SL}_2(\mathbb{C})$ matrices can also be interpreted in terms of their action on a~ste\-reo\-graphic
projection of the complex plane onto the unit sphere using the formula
\begin{gather*}
(X,Y,Z) =\left(\frac{2x}{1+x^2+y^2},\frac{2y}{1+x^2+y^2},\frac{-1+x^2+y^2}{1+x^2+y^2} \right).
\end{gather*}
For real $t_1$, the action of $R_1(t_1)$ is to rotate the roots around the~$X$ axis by an angle $t_1$ counter-clockwise
(i.e., from~$Y$ to~$Z$).
Similarly, the action of $R_2(t_2)$ is to rotate the roots around the~$Y$ axis clockwise (i.e., from~$X$ to~$Z$).
Finally, the action of $R_3(t_3)$ is to rotate the roots around the~$Z$ axis clockwise (i.e., from~$Y$ to~$X$).

The action of these rotations on $\mathbb{R}^3$ can be recovered as
\begin{gather*}
R_1=\left[
\begin{matrix} 1& 0& 0
\\
0& \cos(t_1)& -\sin(t_1)
\\
0&\sin(t_1)&\cos(t_1)
\end{matrix}
\right],
\qquad
R_2=\left[
\begin{matrix} \cos(t_2)& 0& \sin(t_2)
\\
0& 1& 0
\\
-\sin(t_2)&0&\cos(t_2)
\end{matrix}
\right],
\\
R_3=\left[
\begin{matrix}
\cos(t_3)& -\sin(t_3)&0
\\
\sin(t_3)& \cos(t_3)& 0
\\
0&0&1
\end{matrix}
\right].
\end{gather*}
In total, the change of coordinates for the potential is given~by
\begin{gather*}
{\bf y}=cR_3(t_3)R_{2}(t_2)R_{1}(t_1)({\bf x} -{\bf x}_0)+{\bf y}_0.
\end{gather*}
Notice that moving the regular point corresponds to translating the coordinates.

For the remainder of this section, we give the limits between the systems.
The f\/irst few case involve three or fewer roots, or (as in the case of [3111b]) four roots with a~f\/ixed cross ratio.
The limits between these cases can be achieved solely with the action of ${\rm GL}_2({\mathbb C})$, that is, without appealing
to translation the regular point.
For most of these case the limits can be achieved by f\/ixing one root and collapsing the rest together at inf\/inity (or
another prescribed point in $\mathbb{C}^*$).
Several of the limits are elaborated to give a~more complete understanding of the process.

\subsection[{[6] $V_{A}$ to [0] $V_{O}$}]{[6] $\boldsymbol{V_{A}}$ to [0] $\boldsymbol{V_{O}}$}

The covariant polynomial for the $[0]$ equivalence class is simply 0, so to contract down to this
equivalence class, we need only scale by a~factor of $\epsilon^{-1}$ and move the regular point appropriately.
For $V_A$, the regular point is already ${\bf x}_0=(0,0,0)$ so it is unaf\/fected by the scaling.
The contraction is then
\begin{gather*}
%\label{AtoO}
t_1=t_2=t_3=0,
\qquad
c=\epsilon^{-1},
\qquad
{\bf x}_0=(0,0,0),
\qquad
{\bf y}_0=(0,0,0),
\\
q_{A}(0,0,0)(z)=-iz^6 \ \rightarrow \ q_{O}(0,0,0)(z)=0.
\end{gather*}
We start with the potential
\begin{gather*}
V_{A} = a\left(\frac{1}{2}\big({x_1}^2+ {x_2}^2+ {x_3}^2\big)+\frac{1}{12}(x_1-i x_2)^3\right) +b \left(x_1+\frac{1}{4}(x_1-i x_2)^2\right)
\\
\phantom{V_{A} =}{}
+c \left(x_2-\frac{i}{4}(x_1-i x_2)^2\right) +d x_3 +e.
\end{gather*}
The conformal change of coordinates $\mathbf{y}=\epsilon^{-1}\mathbf{x}$ introduces a~factor of $\epsilon^{-2}$ to the
second order terms in the Hamiltonian (due to the conformal scaling of the metric) and so we consider the rescaled
potential $\hat{V}_A=\epsilon^{2} V_A$.
Adjusting the parameters to be
\begin{alignat*}{3}
& \hat{a}=\hat{V}_{A,ee} ({\bf y}_0)=\epsilon^4V_{A,ee}({\bf x}_0)=\epsilon^4 a,
\qquad&&
\hat{b}=\hat{V}_{A,1} ({\bf y}_0)=\epsilon^3V_{A,1}({\bf x}_0)=\epsilon^3 b,&
\\
& \hat{c}=\hat{V}_{A,2} ({\bf y}_0)=\epsilon^3V_{A,2}({\bf x}_0)=\epsilon^3 c,
\qquad &&
\hat{d}=\hat{V}_{A,3} ({\bf y}_0)=\epsilon^3V_{A,3}({\bf x}_0)=\epsilon^3 d,&
\\
& \hat{e}=\hat{V}_{A}({\bf y}_0)=\epsilon^2V_{A}({\bf x}_0)=\epsilon^2 e,&&&
\end{alignat*}
gives the potential
\begin{gather*}
\hat{V}_A=\epsilon^{2} V_A= \hat{a} \left(\frac{1}{2}\big({y_1}^2+ {y_2}^2+ {y_3}^2\big)+\frac{\epsilon}{12}(y_1-i y_2)^3\right)
+\hat{b}\left(y_1+\frac{\epsilon}{4}(y_1-i y_2)^2\right)
\\
\phantom{\hat{V}_A=}{}
+\hat{c} \left(y_2-\frac{i\epsilon}{4}(y_1-i y_2)^2\right) +\hat{d} y_3 +\hat{e},
\end{gather*}
which tends in the $\epsilon\rightarrow 0$ limit to the potential
\begin{gather*}
V_{O}= \frac{\hat{a}}{2} \big(y_1^2+y_2^2+y_3^2\big)+\hat{b} y_1+\hat{c} y_2+\hat{d} y_3+\hat{e}.
\end{gather*}

\subsection[{[51] $V_{\rm VII}$ to [6] $V_{A}$}]{[51] $\boldsymbol{V_{\rm VII}}$ to [6] $\boldsymbol{V_{A}}$}

The regular points for $V_{\rm VII}$ and $V_{A}$ are the origins, ${\bf x}_0=0$ and ${\bf y}_0=0$. The change of variables
is given~by
\begin{gather*}
%\label{7to6}
c=9/2 {\epsilon}^{-3},
\qquad
t_{{1}}=0,
\qquad
t_{{2}}=\frac{\pi}{2},
\qquad
t_{{3}}=i\ln\left(\epsilon\right),
\end{gather*}
which transforms the polynomial
\begin{gather*}
q_{\rm VII}(0,0,0)(z)=-36iz \ \rightarrow \ q_{A}(0,0,0)(z)=-iz^6.
\end{gather*}

Consider f\/irst the roots of the (projective) sextic $q_{\rm VII}(0,0,0)=-36iz$.
There are f\/ive at $z=\infty$ and one at $z=0$.
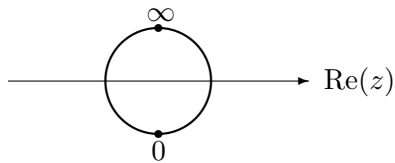
\begin{center}
\setlength{\unitlength}{1mm}
\begin{picture}
(30,20)(-15,-10)\put(-20,0){\vector(2,0){40}}\thicklines \put(0,0){\circle{15}} \put(0,7){\circle*{1}}
\put(0,-7){\circle*{1}} \put(-1.6,8){$\infty$} \put(-0.9,-10.5){$0$} \put(22,-1){$\mathrm{Re}(z)$}
\end{picture}
\setlength{\unitlength}{1pt}
\end{center}
In terms of a~stereographic projection of the complex plane $z=x+iy$ (with the plane of projection intersecting the equator)
the action of $ R_2(t_2)$ is to rotate the sphere around the~$y$-axis (i.e., the imaginary axis which is directed into the plane) by the angle~$t_2$.
So consider
\begin{gather*}
\rho(R_2(\tfrac{\pi}{2})) =\left(
\begin{matrix}
\frac{1}{\sqrt{2}} & -\frac{1}{\sqrt{2}}
\vspace{1mm}\\
\frac{1}{\sqrt{2}} & \frac{1}{\sqrt{2}}
\end{matrix}
\right).
\end{gather*}
Under the action described above the covariant polynomial becomes
\begin{gather*}
\rho(R_2(\tfrac{\pi}{2}))\circ q_{\rm VII}=-\frac{9}{2}i(z+1)^5(z-1),
\end{gather*}
which has brought the f\/ifth root down from inf\/inity down to $z=-1$ and moved the root at zero up to $z=1$.

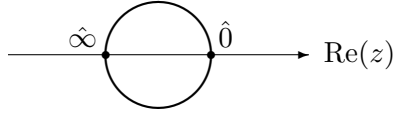
\begin{center}
\setlength{\unitlength}{1mm}
\begin{picture}
(30,20)(-15,-10) \put(-20,0){\vector(2,0){40}}\thicklines \put(0,0){\circle{15}} \put(-7,0){\circle*{1}}
\put(7,0){\circle*{1}} \put(-12,1){$\hat{\infty}$} \put(8,1){$\hat{0}$} \put(22,-1){$\mathrm{Re}(z)$}
\end{picture}
\setlength{\unitlength}{1pt}
\end{center}

Setting $t_3=i\ln(\epsilon)$ gives the matrix
\begin{gather*}
\rho(R_3(i\ln\left(\epsilon\right))) =\left(
\begin{matrix}
\epsilon^{-1/2} & 0
\\
0 & \epsilon^{1/2}
\end{matrix}
\right).
\end{gather*}
On the polynomial this induces the change
\begin{gather*}
\left(\rho(R_2(\tfrac{\pi}{2})) \rho(R_3(i\ln\epsilon))\right)\circ
q_{\rm VII}=-\frac{9}{2}i\epsilon^{-3}(z+\epsilon)^5(z-\epsilon).
\end{gather*}
So f\/inally, scaling by $c=\frac{9}{2}\epsilon^{-3}$ gives
\begin{gather*}
\left(c\rho(R_2(\tfrac{\pi}{2})) \rho(R_3(i\ln\epsilon))\right)\circ q_{\rm VII}=-i(z+\epsilon)^5(z-\epsilon) \ \rightarrow \
q_{A}= -i z^6
\qquad
\text{as}
\quad
\epsilon\rightarrow 0.
\end{gather*}
This limit can be visualized as all the points on the sphere (except~$\infty$) being drawn down towards the origin.
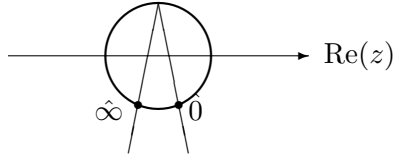
\begin{center}
\setlength{\unitlength}{1mm}
\begin{picture}
(30,30)(-15,-15) \put(-20,0){\vector(2,0){40}}\put(0,7){\line(1,-5){4}} \put(0,7){\line(-1,-5){4}} \thicklines
\put(0,0){\circle{15}} \put(-2.7,-6.5){\circle*{1}} \put(2.7,-6.5){\circle*{1}} \put(-8.5,-8.5){$\hat{\infty}$}
\put(4,-8.5){$\hat{0}$} \put(22,-1){$\mathrm{Re}(z)$}
\end{picture}
\setlength{\unitlength}{1pt}
\end{center}
Or equivalently the roots can remain stationary while the sphere descends into the plane of projection.
From this point of view, the image of any point on the sphere (expect~$\infty$) end up eventually being projected inside
the circle where the plane and sphere intersect, which is shrinking to the origin in the limit.
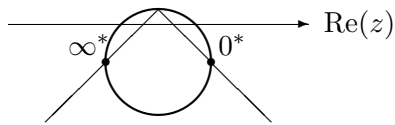
\begin{center}
\setlength{\unitlength}{1mm}
\begin{picture}
(30,20)(-15,-10) \put(-20,5){\vector(2,0){40}}\put(0,7){\line(1,-1){15}} \put(0,7){\line(-1,-1){15}} \thicklines
\put(0,0){\circle{15}} \put(-7,0){\circle*{1}} \put(7,0){\circle*{1}} \put(-12,1){$\infty^*$} \put(8,1){$0^*$}
\put(22,4){$\mathrm{Re}(z)$}
\end{picture}
\setlength{\unitlength}{1pt}
\end{center}

\subsection[{[33] $V_{OO}$ to [6] $V_{A}$}]{[33] $\boldsymbol{V_{OO}}$ to [6] $\boldsymbol{V_{A}}$}

Here, we would like to take the two roots of $q_{OO}$, namely $0$ and~$\infty$, and move them both to~0.
This is going to be essentially the same contraction as the previous one.
The contraction is
\begin{gather*}
%\label{OOtoA}
c=3/4 {\epsilon}^{-3},
\qquad
t_{{1}}=i\ln\left(\epsilon\right),
\qquad
t_{{2}}=-\frac{\pi}{2},
\qquad
t_{{3}}=0.
\end{gather*}
The regular points are
\begin{gather*}
{\bf x}_0=(0,0,1),
\qquad
{\bf y}_0=(0,0,0).
\end{gather*}
The covariant polynomial is transformed as
\begin{gather*}
q_{OO}(0,0,1)(z)=6iz^3 \ \rightarrow \ q_{A}(0,0,0)(z)=-iz^6.
\end{gather*}

\subsection[{[411] $V_{\rm III}$ to [51] $V_{\rm VII}$}]{[411] $\boldsymbol{V_{\rm III}}$ to [51] $\boldsymbol{V_{\rm VII}}$}

The covariant polynomial for $V_{\rm III}$ is given by $q_{\rm III}=-3(1+3z^2)$.
In order to transform this polynomial into $q_{\rm VII}=-36iz$ we need to take one of the f\/inite roots and send it to zero
and the other to inf\/inity.
The contraction that accomplishes this is
\begin{gather*}
%\label{IIItoVII}
c = \frac{-9}{32\sqrt{3}}\epsilon^{-2},
\qquad
t_1 = -\frac{2}{3}\pi,
\qquad
t_2 = \pi,
\qquad
t_3 = -i\ln(\epsilon).
\end{gather*}
The regular points for each system do not move and they are
\begin{gather*}
{\bf x}_0=(0,1,0),
\qquad
{\bf y}_0=(0,0,0).
\end{gather*}
The limit of the polynomial is then
\begin{gather*}
q_{\rm III}(0,1,0)(z)=-9z^2-3 \ \rightarrow \ q_{\rm VII}(0,0,0)(z)=-36iz.
\end{gather*}

\subsection[{[3111b] $V_{\rm VI}$ to [51] $V_{\rm VII}$}]{[3111b] $\boldsymbol{V_{\rm VI}}$ to [51] $\boldsymbol{V_{\rm VII}}$}

The necessary contraction is given~by
\begin{gather*}
%\label{V6toV7}
c=\frac{i}{16}\epsilon^{-1},
\qquad
t_1 =\frac{\pi}{2},
\qquad
t_2 =0,
\qquad
t_3 =-\frac{i}{2}\ln(\epsilon),
\\
{\bf x}_0=(0,0,2i),
\qquad
{\bf y}_0=(0,0,0).
\end{gather*}
In this limit, all of the roots coalesce to inf\/inity except for a~single root at zero.
The limit of the polynomial is then
\begin{gather*}
q_{\rm VI}(0,0,2i)(z)=3iz^6+3z^3 \ \rightarrow \ q_{\rm VII}(0,0,0)(z)=-36iz.
\end{gather*}

\subsection[{[3111b] $V_{\rm VI}$ to [33] $V_{OO}$}]{[3111b] $\boldsymbol{V_{\rm VI}}$ to [33] $\boldsymbol{V_{OO}}$}

For this contraction, we would like to move the roots of the polynomial $q_{\rm VI}$ by sending the
three non-zero roots to inf\/inity and leaving the 0 root f\/ixed.
This can be accomplished by choosing an imaginary, singular value for $t_3$ which has the ef\/fect of scaling the
variable~$z$.
The required contraction is then
\begin{gather*}
%\label{V6toVOO}
t_3=-i\ln (\epsilon),
\qquad
c=\frac 12,
\qquad
{\bf x}_0=(0,0,2),
\qquad
{\bf y}_0=(0,0,1).
\end{gather*}
The polynomials transform as
\begin{gather*}
q_{\rm VI}(0,0,2)(z)=3iz^3\big(1+z^3\big) \ \rightarrow \ q_{OO}(0,0,1)(z)=6iz^3.
\end{gather*}

\subsection[{[3111a] $V_{\rm II}$ to [411] $V_{\rm V}$}]{[3111a] $\boldsymbol{V_{\rm II}}$ to [411] $\boldsymbol{V_{\rm V}}$}

From this case onward the polynomials involve four or more roots and require additional information
about the cross ratios of the roots in order to dif\/ferentiate between equivalence classes.
Translation of the regular point allows the cross ratios of these roots to change.
Without translation of the regular point the only way singular limits could be taken is by collapsing all, save one, of
the roots together.
This means that, starting at a~point which has represents root multiplicities [3111] the only possible limits without
translation of the regular point are [51],[33] or their degenerations.
In all of the examples that follow, translation of the regular point will be necessary to modify the cross ratios (i.e.,
move the roots in a~way which the M\"obius transformations cannot) and, in ef\/fect, rescue some of the roots in the singular limit.

To contract the covariant polynomial of $V_{\rm II}$, we observe that the polynomial evaluated at $(x_1,x_2,x_3)$ is
\begin{gather*}
q_{\rm II}(x_1, x_2, x_3)(z)=3i\left(\frac{2}{x_3}z^3+\frac{3}{x_1+i x_2}z^2+\frac{x_1-i x_2}{(x_1+i x_2)^2}\right).
\end{gather*}
Specializing the f\/irst two coordinates to $(x_1,x_2)=(0,-i)$ gives
\begin{gather*}
q_{\rm II}(0, -i, x_3)(z)=3i\left(\frac{2}{x_3}z^3+3z^2-1\right),
\end{gather*}
and so by moving the regular point to~$\infty$, we obtain the desired contraction.
The regular points and scaling parameter is then chosen as
\begin{gather*}
%\label{V2toV3}
c=i,
\qquad
{\bf x}_0=\big(0, -i, \epsilon^{-1}\big),
\qquad
{\bf y}_0=(0,1,0).
\end{gather*}
This contraction gives the required limits of the polynomial,
\begin{gather*}
q_{\rm II}(0, -i, 1)(z)=3i\big(2z^3+3z^2-1\big) \ \rightarrow \ q_{\rm V}(0,1,0)(z)=9z^2-3.
\end{gather*}

\subsection[{[3111a] $V_{\rm II}$ to [3111b] $V_{\rm VI}$}]{[3111a] $\boldsymbol{V_{\rm II}}$ to [3111b] $\boldsymbol{V_{\rm VI}}$}

There are two distinct equivalence classes of systems whose covariant polynomials have the root
structure $[311]$; these equivalence classes are distinguished by the characteristic that the cross ratio of the root be
$\exp(i \pi/3)$ (case [3111b]) or not (case [311a])~\cite{CapelThesis, CapelKress2014}.
For the limit between these two systems, we begin with a~polynomial with a~triple root and three simple roots with
a~cross ratio of the roots away from 0, 1,~$\infty$ or $\exp(\pm i \pi/3)$. We would like to f\/ind a~limit that will
contract this polynomial to one with the same root structure but with a~cross ratio of $\exp(i \pi/3)$. We begin with
the covariant polynomial for $V_{\rm II}$ which has a~triple root at inf\/inity and move this root to 0 by rotating around the
$x_2$ axis by $\pi$. We then move the $x_1$ coordinate of the regular point to inf\/inity and perform a~singular rotation
about $x_3$ via
\begin{gather*}
%\label{V2toV6}
t_2=\pi,
\qquad
t_3=\frac{i}{3}\ln(\epsilon),
\qquad
{\bf x}_0=\big(\epsilon^{-1}, 0, 3\epsilon-2\big),
\qquad
{\bf y}_0=(0,0,2).
\end{gather*}
This limit transforms the polynomials as (from $\epsilon=1$ to $\epsilon=0$)
\begin{gather*}
q_{\rm II}(1, 0, 1)(z)=3i\big(1+3z^2+2{z^3}\big) \ \rightarrow \ q_{\rm VI}(0,0,2)(z)=3iz^3\big(z^3+1\big).
\end{gather*}

\subsection[{[111111c] $V_{\rm IV}$ to [411] $V_{\rm V}$}]{[111111c] $\boldsymbol{V_{\rm IV}}$ to [411] $\boldsymbol{V_{\rm V}}$}

For this contraction, we take the equivalence class described by a~polynomial with 6 simple roots
satisfying the multi-ratio condition, a~discriminating condition for the [111111] root structu\-re~\mbox{\cite{CapelThesis,
CapelKress2014}}, to one with two simple f\/inite roots and the remaining at inf\/inity.
The coordinate limit is given~by
\begin{gather*}
%\label{V4toV5}
c=\frac{-1}{2\epsilon},
\qquad
t_1=0,
\qquad
t_2=-\frac{\pi}{2},
\qquad
t_3=-i\ln (\epsilon).
\\
{\bf x}_0=\left(0, \frac{1}{\epsilon^2-1}, \frac{i}{\epsilon^2+1}\right),
\qquad
{\bf y_0} =(0,1,0).
\end{gather*}
The covariant polynomial contracts as
\begin{gather*}
q_{\rm IV}(x_1, x_2, x_3)(z)=\frac{6iz^3}{x_3}+\frac{3(z^2+1)^3}{4x_2} \ \rightarrow \ q_{\rm V}(0,1,0)(z)=9z^2-3.
\end{gather*}

\subsection[{[111111c] $V_{\rm IV}$ to [3111b] $V_{\rm VI}$}]{[111111c] $\boldsymbol{V_{\rm IV}}$ to [3111b] $\boldsymbol{V_{\rm VI}}$}

As seen above, the covariant polynomial for the potential $V_{\rm IV}$ is given~by
\begin{gather*}
q_{\rm IV}(x_1, x_2, x_3)(z)=\frac{6i}{x_3} z^3+\frac{3}{4x_2}\big(z^2+1\big)^3.
\end{gather*}
An obvious easy way to achieve a~triple root is to let the second coordinate $x_2$ tend to inf\/inity.
However this has the unintended side ef\/fect of collapsing the other three roots at inf\/inity.
The singular rotation $R_3(i\ln(\epsilon))$ can be used to counteract this growth, provided the parameter~$\epsilon$
correctly balances the growth of $x_2$.
This action only serves to accelerate the roots tending to zero, and hence gives us the limit we require.

This contraction is accomplished by a~singular rotation in around the $x_3$-axis and a~singular translation of the regular point.
The contraction is
\begin{gather*}
%\label{VIVtoVVI}
t_3=i\ln \epsilon,
\qquad
{\bf x}_0=\left(0, \frac{-i}{4\epsilon^3}, 2\right),
\qquad
{\bf y}_0=(0,0,2).
\end{gather*}
The covariant polynomials contract as
\begin{gather*}
q_{\rm IV}(x_1, x_2, x_3)(z)=\frac{6iz^3}{x_3}+\frac{3(z^2+1)^3}{4x_2} \ \rightarrow \ q_{\rm VI}(0,0,2)(z)=3iz^3\big(z^3+1\big).
\end{gather*}

\subsection[{[111111b] $V_{\rm I}$ to [3111a] $V_{\rm II}$}]{[111111b] $\boldsymbol{V_{\rm I}}$ to [3111a] $\boldsymbol{V_{\rm II}}$}

Here we would like to send 3 of the f\/inite roots to inf\/inity.
The covariant polynomial for $V_{\rm I}$ is
\begin{gather*}
q_{\rm I}(x_1,x_2, x_3)(z)={\frac {3 \left(x_{{1}}+ix_{{2}} \right) {z}^{6}}{4x_{{1}}x_{{2}}}}+{\frac
{9\left(x_{{1}}-ix_{{2}} \right) {z}^{4}}{4x_{{1}}x_{{2}}}}+{\frac {6 i{z}^{3}}{x_{{3}}}}+{\frac
{9\left(x_{{1}}+ix_{{2}} \right) {z}^{2}}{4x_{{1}}x_{{2}}}}+{\frac {3(x_{{1}}-ix_{{2}})}{4 x_{{1}}x_{{2}}}}\!
\\
\phantom{q_{\rm I}(x_1,x_2, x_3)(z)}
 = \frac{3i}{4x_1}\big(z^2-1\big)^3 +\frac{3}{4x_2}\big(z^2+1\big)^3 +\frac{6i}{x_3}z^3.
\end{gather*}
Rotations around the~$z$ axis have the ef\/fect of multiplying the variable~$z$ by $e^{-2it_3}$ and the entire polynomial
by a~factor of $e^{6it_3}$.
Thus if we take a~purely imaginary angle, the $z^3$ variable will be unchanged and it will be possible to send the
coef\/f\/icient of $z^4$ and $z^6$ to 0.
In order to retain the lower order terms, it is necessary to move the regular point so as to of\/fset the singular limit.
The required contraction is
\begin{gather*}
%\label{VItoV2}
t_3=-i\ln \epsilon,
\qquad
{\bf x}_0=\left({\frac {{\epsilon}^{2}+1}{2\epsilon}},{\frac {i \left({\epsilon}^{2}-1 \right)}{2\epsilon}},1\right),
\qquad
{\bf y}_0=(1,0,1).
\end{gather*}
The polynomials transform as
\begin{gather*}
q_{\rm I}(x_1,x_2, x_3)(z)= \frac{3i}{4x_1}\big(z^2-1\big)^3 +\frac{3}{4x_2}\big(z^2+1\big)^3 +\frac{6i}{x_3}z^3 \quad \rightarrow
\\
q_{\rm II}(1,0,1)(z)=3 i \big(1+2 {z}^{3}+3 {z}^{2} \big).
\end{gather*}

\subsubsection{``Geometric inspired limit''}

This contraction can be carried out in a~manner similar to the previous one.
By moving the regular point correctly we can collapse three roots at $z=0$ and three roots are $z=\infty$.
Using the same singular rotation as before, the collapse of one of these triplets of roots can halted, and the other
accelerated.

The covariant polynomial for $V_{\rm I}$ is given~by
\begin{gather*}
q_{\rm I}(x_1, x_2, x_3)(z)  = \frac{3i}{4x_1}\big(z^2-1\big)^3 +\frac{3}{4x_2}\big(z^2+1\big)^3 +\frac{6i}{x_3}z^3.
\end{gather*}
By choosing to work on the hypersurface $x_2=-ix_1$ we ensure that $q_{\rm I}$ has a~double root at $z=0$.
The choice $x_1=\eta$, $x=-i\eta$ gives
\begin{gather*}
q_{\rm I}(\eta, -i\eta, x_3)(z) = \frac{3i}{2}z^2\left(\frac{z^4}{\eta}+\frac{4z}{x_3}+\frac{3}{\eta}\right).
\end{gather*}
Sending~$\eta$ to inf\/inity gives the limit collapses a~single root to join the double root at zero, and sends the other
three roots to inf\/inity.
Using a~singular rotation around the $x_3$ axis the root tending to zero can be saved, and the roots tending inf\/inity accelerated.

The required contraction is therefore
\begin{gather*}
%\label{VItoV2geom}
c=\frac12,
\qquad
t_3=-i\ln \epsilon,
\qquad
{\bf x}_0=\big(\epsilon^{-1},-i\epsilon^{-1},4\big),
\qquad
{\bf y}_0=\big(\tfrac{1}{2},\tfrac{-i}{2},2\big).
\end{gather*}
The polynomials transform as
\begin{gather*}
q_{\rm I}(x_1, x_2, x_3)(z)= \frac{3i}{4x_1}\big(z^2-1\big)^3 +\frac{3}{4x_2}\big(z^2+1\big)^3 +\frac{6i}{x_3}z^3
\quad
\rightarrow
\\
q_{\rm II}\left(\frac12, \frac{-i}{2}, 2\right)(z)=3 i z^3+9 iz^2.
\end{gather*}

\subsection[{[111111b] $V_{\rm I}$ to [111111c] $V_{\rm IV}$}]{[111111b] $\boldsymbol{V_{\rm I}}$ to [111111c] $\boldsymbol{V_{\rm IV}}$}

To obtain this contraction, we move the regular point and scale the potential.
The required contraction is
\begin{gather*}
%\label{V1toV4}
c=\frac{1}{\epsilon},
\qquad
{\bf x}_0=(1,\epsilon, \epsilon),
\qquad
{\bf y}_0=(0,1,1).
\end{gather*}
This transforms the covariant polynomials as
\begin{gather*}
q_{\rm I}(x_1, x_2, x_3)(z)= \frac{3i}{4x_1}\big(z^2-1\big)^3 +\frac{3}{4x_2}\big(z^2+1\big)^3 +\frac{6i}{x_3}z^3 \quad \rightarrow\\
q_{\rm IV}(0,1,1)(z)=6iz^3+\frac{3}{4}\big(z^2+1\big)^3.
\end{gather*}

\subsection[{[111111a] $V_{\rm SW}$ to [111111b] $V_{\rm I}$}]{[111111a] $\boldsymbol{V_{\rm SW}}$ to [111111b] $\boldsymbol{V_{\rm I}}$}

The case [111111a] covers the most generic forms for the covariant polynomials.
A~particularly nice example of the polynomials in this class is
\begin{gather*}
q_{\rm SW}({\bf x}_0)(z)=\frac{9}{2} z(z-1)(z+1)(z-i)(z+i),
\qquad
{\bf x}_0=(1,1,1).
%\label{SWpoly}
\end{gather*}
However, this polynomial doesn't show the full regular point dependence of $q_{\rm SW}(z)$

One possible contraction is given by a~scaling by $\epsilon^{-1}$ and translation of the regular point to the origin
\begin{gather*}
c=\frac1{\epsilon},
\qquad
{\bf x}_0=(\epsilon, \epsilon, \epsilon),
\qquad
{\bf y}_0=(1,1,1).
\end{gather*}
Following the regular point $(\epsilon, \epsilon, \epsilon)$ the covariant polynomial for the is given~by
\begin{gather*}
q_{\rm SW}(\epsilon, \epsilon, \epsilon)(z)
=\frac{3i(1-\epsilon^4)(1+i)}{4(1-9\epsilon^4)}+\frac{36 \epsilon^4 z}{(1-9\epsilon^4)}
-\frac{9(1-\epsilon^4)(1+i) z^2}{4 (1-9\epsilon^4)}-\frac{6i (1-\epsilon^4) z^3}{(1-9\epsilon^4)}\\
\phantom{q_{\rm SW}(\epsilon, \epsilon, \epsilon)(z)}{}
+\frac{9i(1-\epsilon^4)(1+i) z^4}{4(1-9\epsilon^4)}
-\frac{36\epsilon^4 z^5}{(1-9\epsilon^4)}-\frac{3(1-\epsilon^4)(1+i) z^6}{4(1-9\epsilon^4)}.
\end{gather*}

Here~$\epsilon$ must follow a~path from 1 to 0 in the complex plane, avoiding points where $\epsilon^4=\frac19$.
The limit of this contraction transforms the covariant polynomial to
\begin{gather*}
q_{\rm I}(1,1,1)(z)= \frac{3i}{4}\big(z^2-1\big)^3 +\frac{3}{4}\big(z^2+1\big)^3 +6iz^3.
\end{gather*}

\section{Conclusion}\label{Section4}

By using the invariant classif\/ication of conformally superintegrable systems on conformally f\/lat 3D manifolds, we have
been able to give explicitly the coordinate limits between the 10 equivalence classes.
We have therefore proven that all such systems are limits of the generic system,~$V_{\rm SW}$.
We would like to point out that this analysis relied on the potentials begin complex, beginning with the fact that we
are determining only equivalence classes under the St\"ackel transform which does not necessarily preserve real forms
and continuing into the classif\/ication using the action of the conformal group.
It would therefore be interesting to understand the classif\/ication of the real forms of these potentials and their
corresponding limits, which would be non-trivial applications of the complex theory.

One of the most interesting avenues of this research is the connection between superintegrable systems, orthogonal
polynomials (OPs) and special functions.
From the beginning of the study of superintegrable systems, going back to the harmonic oscillator, Kepler Coulomb system
and the discovery in '65 of multiseparable systems by Smorodinsky, Winternitz and collaborators~\cite{FMSUW,MSVW, FSUW}, there has been a~clear connection between superintegrable systems and the special functions of mathematical
physics.
In particular, the wave functions of superintegrable systems have been expressed in terms of OPs~\cite{KMJP} and all
superintegrable systems are conjectured to have this exactly-solvable nature~\cite{TempTW}.
More recently, this connection has been extended by studying the representation theory of the symmetry algebras
associated with superintegrable systems and interbasis expansion coef\/f\/icients~\cite{genest2014generic, genest2014interbasis, genest2013superintegrability, Zhedanov1992a, Zhedanov1992b,
KMPost3, KMPost10, POST2011}.
In~\cite{KMPost13}, the Askey scheme of classical hypergeometric OPs in one variable as well as their limits were
related to superintegrable systems in 2D and the contractions of their symmetry algebras.
Work is currently ongoing to apply the limits obtained in this paper to f\/ind contractions of the associated quadratic
algebras and their polynomial representations.

Finally, we would like to mention that as is evident from the classif\/ication theory, the superintegrable systems are
uniquely determined by the functionally linearly independent set of second-order Killing tensors.
Therefore we expect, as in the 2D case~\cite{KMContractions2014}, that the quadratic algebras and their contractions
will be completely determined by contractions of the underlying Lie algebra.
However, this is still an open question and one which we anticipate will be aided by understanding the connection
between the invariant classif\/ication and the quadratic algebras.

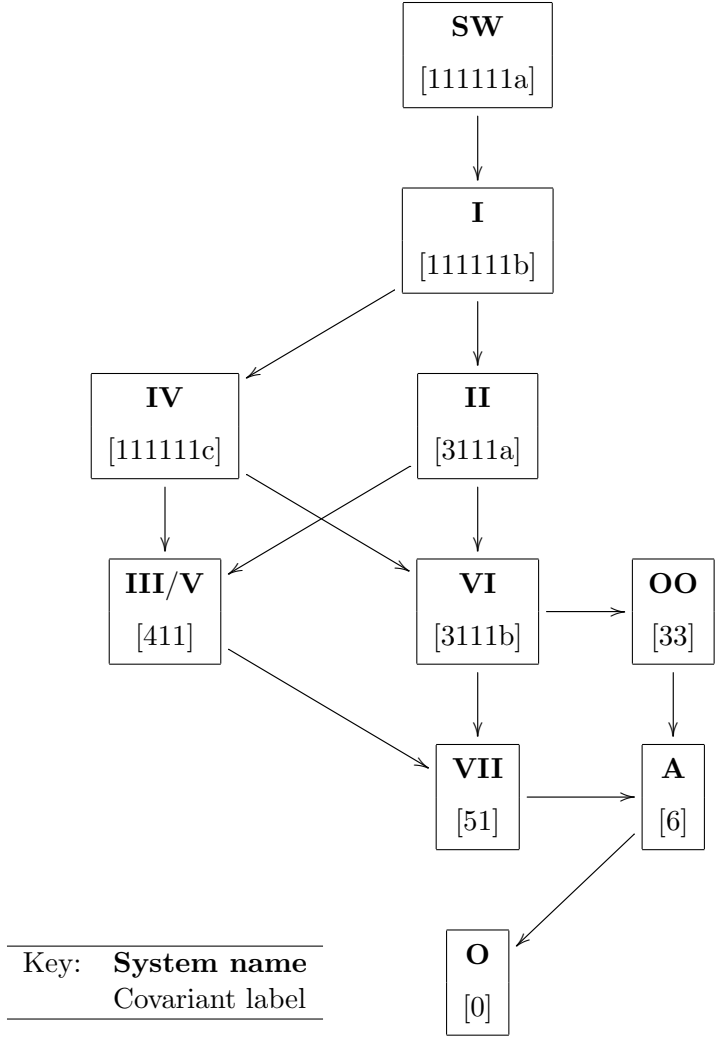
\begin{figure}[t!] \centering
\begin{gather*}
\xymatrix{
  & { \begin{tabular}{|c|} \hline $\vphantom{\Big|}\mathbf{SW}$ \\ $\vphantom{\Big|}$[111111a] \\ \hline \end{tabular}} \ar[d]
  &
 \\ %%%%%%%%%%%%%%%%%%%%%%%
  & { \begin{tabular}{|c|} \hline $\vphantom{\Big|}\mathbf{I}$ \\ $\vphantom{\Big|}$[111111b] \\ \hline \end{tabular}}  \ar[d]\ar[dl]
  &
 \\ %%%%%%%%%%%%%%%%%%%%%%%
  { \begin{tabular}{|c|} \hline $\vphantom{\Big|}\mathbf{IV}$ \\ $\vphantom{\Big|}$[111111c]\\ \hline \end{tabular}}\ar[dr]\ar[d]
  & { \begin{tabular}{|c|} \hline $\vphantom{\Big|}\mathbf{II}$  \\ $\vphantom{\Big|}$[3111a] \\ \hline \end{tabular}} \ar[d]\ar[dl]
  &
 \\ %%%%%%%%%%%%%%%%%%%%%%%
  { \begin{tabular}{|c|} \hline {$\vphantom{\Big|}\mathbf{III/V}$}\\ $\vphantom{\Big|}$[411] \\ \hline \end{tabular}}\ar[dr]
  & { \begin{tabular}{|c|} \hline $\vphantom{\Big|}\mathbf{VI} $ \\ $\vphantom{\Big|}$[3111b] \\ \hline \end{tabular}}\ar[d]\ar[r]
  & { \begin{tabular}{|c|} \hline $\vphantom{\Big|}\mathbf{OO}$\\ $\vphantom{\Big|}$[33] \\ \hline \end{tabular}}\ar[d]
 \\ %%%%%%%%%%%%%%%%%%%%%%%
& { \begin{tabular}{|c|} \hline $\vphantom{\Big|}\mathbf{VII}$ \\ $\vphantom{\Big|}$[51] \\ \hline \end{tabular}} \ar[r]
& { \begin{tabular}{|c|} \hline $\vphantom{\Big|}\mathbf{A}$ \\ $\vphantom{\Big|}$[6] \\ \hline \end{tabular}} \ar[dl]
 \\ %%%%%%%%%%%%%%%%%%%%%%%
 { \begin{tabular}{ rl } \hline Key:& \textbf{System name} \\ & Covariant label \\ \hline \end{tabular}}
& { \begin{tabular}{|c|} \hline $\vphantom{\Big|}\mathbf{O}$ \\  $\vphantom{\Big|}$[0] \\ \hline \end{tabular}}
&
}
\end{gather*}
\caption{Limiting diagram.}
%\label{figure:detailed_ideal_containment_diagram}
\end{figure}

\appendix

\section{Potentials}

Here we list the potentials discussed above.
Each of the limits above has been chosen so that the limit potential is in the form given below.
Note that the parameters are chosen is for aesthetic reasons, to give the potentials in the most simple form.
They are related by a~linear transformation to
the parameters $\{V ({\bf x}_0), V_1 ({\bf x}_0), V_2 ({\bf x}_0), V_3({\bf x}_0), V_{ee} ({\bf x}_0)\}$,
which of course depend on the regular point ${\bf x}_0$.

Note that $V_{\rm V}$ is St\"ackel equivalent to $V_{\rm III}$ by choice of the St\"ackel multiplier $U=(x_1+i x_2)^{-2}$:
\begin{gather*}
V_O= \frac{a}{2} \big(x_1^2+x_2^2+x_3^2\big)+b x_1+c x_2+d x_3+e,
\\
V_{A} = a\left(\frac{1}{2}\big({x_1}^2+ {x_2}^2+ {x_3}^2\big)+\frac{1}{12}(x_1-i x_2)^3\right) +b \left(x_1+\frac{1}{4}(x_1-i x_2)^2\right)
\\
\phantom{V_{A} =}{}
+c \left(x_2-\frac{i}{4}(x_1-i x_2)^2\right) +d x_3 +e,
\\
V_{OO} =a\big(4 x_1^2+4 x_2^2+x_3^2\big)+b x_1+c x_2+\frac{d}{x_3^2}+e,
\\
V_{\rm VII} = a(x_1+i x_2)+b \left(3 (x_1+i x_2)^2+x_3\right)
\\
\phantom{V_{\rm VII} =}{}
+c\left(16 (x_1+i x_2)^3+(x_1-i x_2)+12 x_3 (x_1+i x_2)\right)
\\
\phantom{V_{\rm VII} =}{}
+d\big(5 (x_1+i x_2)^4+
x_1^2+x_2^2+x_3^2\big)+6\big(
(x_1+i x_2)^2 x_3\big)+e,
\\
V_{\rm VI} = a\big(x_3^2-2 (x_1-i x_2)^3+4 \big(x_1^2+x_2^2\big)\big)+b \big(2 x_1+2 i x_2-3 (x_1-i x_2)^2\big)
\\
\phantom{V_{\rm VI} =}{}
+c  (x_1-i x_2 )+\frac{d}{x_3^2}+e,
\\
V_{\rm V}=a \big(x_1^2+x_2^2+4 x_3^2\big)+b x_3+\frac{c}{(x_1+i x_2)^2}+d \frac{x_1-i x_2}{(x_1+i x_2)^3}+e,
\\
V_{\rm IV} = a\big(4 x_1^2+x_2^2+x_3^2\big)+b x_1+\frac{c}{x_2^2}+\frac{d}{x_3^2}+e,
\\
V_{\rm III}=a \big(x_1^2+x_2^2+x_3^2\big)+\frac{b}{(x_1+i x_2)^2} + \frac{c x_3}{(x_1+i x_2)^3}+d \frac{x_1^2+x_2^2-3 x_3^2}{(x_1+i x_2)^4}+e,
\\
V_{\rm II} = a\big({x_1}^2+{x_2}^2+{x_3}^2\big)+b\frac{(x_1 - i x_2)}{(x_1 + i x_2)^3}
+c\frac{1}{(x_1 + i x_2)^2}+d\frac{1}{{x_3}^2}+e,
\\
V_{\rm I}=a\big(x_1^2+x_2^2+x_3^2\big)+\frac{b}{x_1^2}+\frac{c}{x_2^2}+\frac{d}{x_3^2}+e,
\\
V_{\rm SW}=\frac{a}{\big(1+x_1^2+x_2^2+x_3^2\big)^2}+\frac{b}{x_1^2}+\frac{c}{x_2^2}+\frac{d}{x_3^2}+\frac{e}{\big({-}1+x_1^2+x_2^2+x_3^2\big)^2}.
\end{gather*}

\pdfbookmark[1]{References}{ref}
\LastPageEnding

\end{document}